\documentclass[oribibl]{llncs}

\usepackage[english]{babel}
\usepackage[utf8x]{inputenc}
\usepackage{amsmath}
\usepackage{graphicx}
\usepackage{caption}
\usepackage{url}
\usepackage{epstopdf}
\usepackage[colorinlistoftodos]{todonotes}
\usepackage{times}
\usepackage{array}

\usepackage{color}
\definecolor{gray}{rgb}{0.4,0.4,0.4}
\definecolor{darkblue}{rgb}{0.0,0.0,0.6}
\definecolor{cyan}{rgb}{0.0,0.6,0.6}
\usepackage{listings}
\lstdefinelanguage{XML}
{
  morestring=[b]",
  morestring=[s]{>}{<},
  morecomment=[s]{<?}{?>},
  stringstyle=\color{black},
  identifierstyle=\color{darkblue},
  keywordstyle=\color{cyan},
  morekeywords={xmlns,version,type}
}

\lstdefinelanguage{Black}
{
  stringstyle=\color{black},
  identifierstyle=\color{darkblue},
  keywordstyle=\color{cyan}
}

\begin{document}

\mainmatter
\title{How to perform research in Hadoop environment not losing mental equilibrium - case study}
\titlerunning{Author disambiguation in the YADDA2 software platform}

\author{Piotr Jan Dendek \and Artur Czeczko \and Mateusz Fedoryszak \and\\ Adam Kawa \and Piotr Wendykier \and Lukasz Bolikowski}
\authorrunning{Piotr Jan Dendek et al.} 
\institute{Interdisciplinary Centre for Mathematical and Computational Modelling,\\ University of Warsaw\\
\email{\{p.dendek, a.czeczko, m.fedoryszak, p.wendykier, l.bolikowski\}@icm.edu.pl, \\
kawa.adam@gmail.com}}

\maketitle  

\begin{abstract}
Conducting a research in an efficient, repetitive, evaluable, but also convenient (in terms of development) way has always been a challenge. To satisfy those requirements in a long term and simultaneously minimize costs of the software engineering process, one has to follow a certain set of guidelines. This article describes such guidelines based on the research environment called Content Analysis System (CoAnSys) created in the Center for Open Science (CeON). Best practices and tools for working in the Apache Hadoop environment, as well as the process of establishing these rules are portrayed.
\end{abstract}


\section{Introduction}
\label{sec:introduction}
\subsection{Distributed Computing}
\label{sec:distributed_computing}

Currently, no single machine can be employed to consume, in a convenient way and in a reasonable time, the massive amounts of data. To tackle this issue, several approaches to GRID computing have been introduced.

In 2004, Google proposed MapReduce paradigm \cite{DeanG:2004,Dean2010}, which gives an opportunity to maximize computer resources utilization and enables better parallelization trade-off between the number of machines used and a computational time. Many articles discussed the topic of embracing MapReduce for data analysis purposes, covering also its optimization and adaptation \cite{Dean2010a,Ranger2007,Yang2007,Zaharia2008,McKenna2010, Chu2007, Kawa2013}. MapReduce is a great candidate to face challenges brought by the universe of big data.

Since 2008, the second name of MapReduce is its open-source, Java based implementation of Apache Hadoop. It has proved its adaptability for business purposes in companies like Amazon, Last.fm, LinkedIn, Twitter, Mendeley, and Yahoo!, e.g. for rapid storing and retrieving huge amounts of data, machine learning and information retrieval mechanisms.

Authors believe that all readers of this article have a basic knowledge about MapReduce and Apache Hadoop environment. In case of doubts we encourage to familiarize with adequate literature \cite{White:2009:HDG:1717298,hbase2011george,Gates2011}.

\subsection{Research Environment}
\label{sec:research_environment}

Thanks to services like Amazon Elastic Compute Cloud, use of the powerful environment of Apache Hadoop is not accompanied with a burden of purchasing appropriate 
machines, followed by onerous administrative tasks. Nevertheless, there are many other difficulties, especially in conducting research, which should be anticipated and resolved before they arise in major impediments.

To omit such occasions, researchers should put on the first place a certain dojo \footnote{\url{http://arkitus.com/PRML/}}, which can be condensed as follows:
\begin{enumerate}
\item Separate input data, working data and output data and keep them for a later comparison (e.g. for other researchers or in case of detecting malformation in a code). Save everything to disk frequently. [\textbf{comparable}, \textbf{convenient}]
\item Separate options from parameters. Make it easy to swap options.[\textbf{convenient}]
\item Modify input data with care. [\textbf{comparable}]
\item Record options used to generate each run of the algorithm. [\textbf{comparable}]
\item Make it easy to execute only portions of the code. [\textbf{convenient}]
\item Trigger all workflows in a fully automated manner. [\textbf{convenient}, \textbf{efficient}]
\end{enumerate}

Naturally, obtaining and shipping research results is the top priority goal - paraphrasing a popular quote “The technology you use impresses no one. The experience you create with it is everything”. To do so, a suitable background and a daily routine are indispensable. Intuitively, following all those rules is equivalent to a time overhead in a research process, especially at the beginning, but later on, it is rewarded by results acceleration and convenience. Moreover, some desperate rescue actions in the final stage of writing articles, typically close to the submission deadline, are no longer needed. Those requirements are not only put on particular persons, but also on the whole research system.

The rest of this paper is organized as follows. Section 2 introduces CoAnSys - a framework for mining scientific publications using Apache Hadoop, Section 3 describes  tools and practices used for the development of CoAnSys. Section 4 describes Chrum - a tool developed by the authors to simplify the description of a workflow. Finally, Section 5 derives conclusions about developing data processing environment and conducting research in Apache Hadoop. 

\section{CoAnSys}
\label{sec:conasys}

\begin{figure} 
	\begin{minipage}[t]{0.64\textwidth}
		\begin{center} 
			\includegraphics[width=\textwidth]{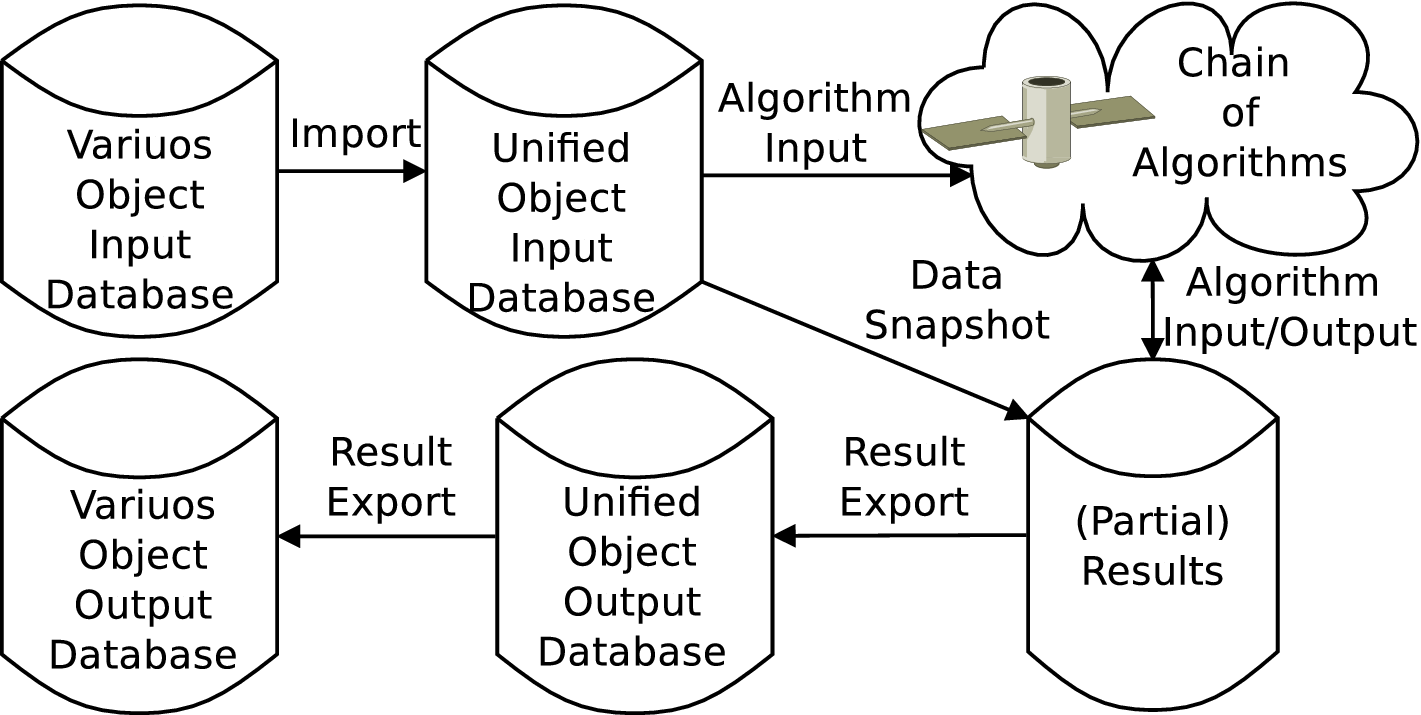}
			\caption{\label{coansys_framework} A generic framework for performing unified, repetitive experiments on data with eventual algorithm comparison and evaluation.} 
		\end{center} 
	\end{minipage} 
\hfill 
	\begin{minipage}[t]{0.35\textwidth}
		\begin{center} 
			\includegraphics[width=\textwidth]{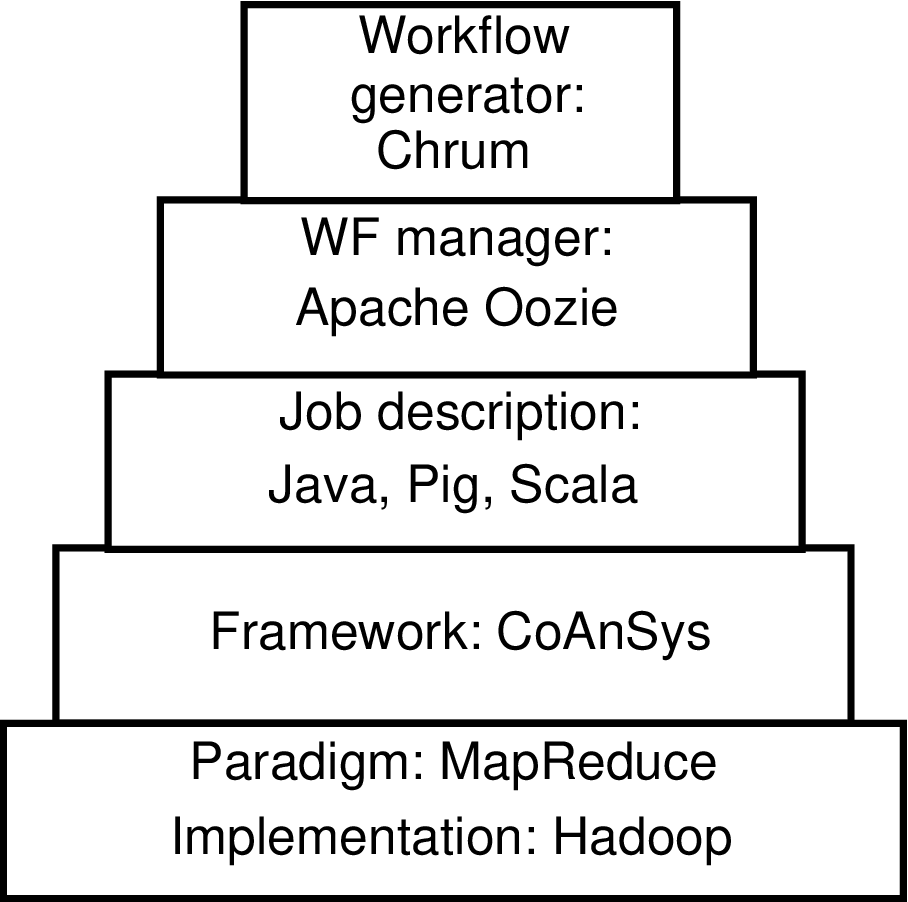}
			\caption{\label{ceon_stack}CeON Analytic Stack.}
		\end{center} 
	\end{minipage} 
\end{figure}

Till now many systems adapting the rules described in Section \ref{sec:research_environment} where introduced, e.g. UIMA \cite{Ferrucci2004}, Synat \cite{Bembenik:1503879}, OpenAIRE \cite{Manghi2010} or currently developed OpenAIREplus \cite{Manghi2012}. Complementary, the Center for Open Science (CeON, part of Interdisciplinary Center for Mathematical and Computational Modeling, University of Warsaw) has been developing, the Content Analysis System (CoAnSys) \footnote{\url{http://github.com/CeON/CoAnSys}} embracing Apache Hadoop ecosystem. This framework is suitable for research on big data, e.g. document metadata extraction and further analysis. An architecture  overview of CoAnSys is illustrated in  Fig. \ref{coansys_framework}. It consists of the following parts:

\begin{enumerate}
\item Data import mechanisms.
\item Data processing mechanisms by a sequential or parallel set of algorithms.
\item Data export mechanisms.
\end{enumerate}

To provide unified input data, CoAnSys utilizes Protocol Buffers \footnote{\url{http://code.google.com/p/protobuf/}}, a method of serializing data into a compact binary format. We have chosen Protocol Buffers over similar Apache Trift \footnote{\url{http://thrift.apache.org/}} or Apache Avro \footnote{\url{http://avro.apache.org/}} due to the fact that it had already established position and was widely used at Google, Twitter and Mendeley. Once the data is serialized, it can be imported into the HBase using REST protocol (HBase has a build-in REST server called Stargate). REST allows for simultaneous import of data from multiple clients, however, to perform complex, offline operations on the data, it is beneficial to use HDFS instead of HBase. 

For that reasons, the next step in the CoAnSys workflow is to copy data from an HBase table to an HDFS sequence file. Data in that form (Unified Object Input Database in Fig. \ref{coansys_framework}) are then used as input for the algorithms. MapReduce jobs are implemented in Java, Apache Pig \cite{Gates2011} or Scala (depends on the preference of a developer) and chained together using Apache Oozie \footnote{\url{http://oozie.apache.org/}} - a workflow scheduler system for Hadoop. To maintain the consistency of the framework, each module has well defined I/O interfaces in the form of Protocol Buffers schemas. This means, that each module reads input data from an HDFS sequence file (that contains records serialized with Protocol Buffers) and stores the results in the same way (output records can have different schemas). In the last part of the workflow, the output data from each module is stored in an HBase table and can be accessed by the clients via REST protocol. From the high-level architecture perspective, CoAnSys workflow is quite similar to UIMA’s collection processing engine, where user defined functions play a role of nodes in the chain of execution.    
   
There are three ongoing projects that use the elements of CoAnSys framework. OpenAIREplus \footnote{\url{http://www.openaire.eu/}} is the European open access data infrastructure for scholarly and scientific communication. PBN \footnote{\url{http://pbn.nauka.gov.pl/}} is a portal of the Polish Ministry of Science and Higher Education, collecting information on publications of Polish scientists and on Polish and foreign scholarly journals. Infona \footnote{\url{http://www.infona.pl/}} is a user portal for SYNAT - Polish national strategic research program to build an interdisciplinary system for interactive scientific information.

Research conducted in CeON may be analyzed in five layers depicted in Fig. \ref{ceon_stack}. Since its very beginning, the development of CoAnSys has been a great lesson to CeON team. We investigated both technology and ours limitations, applied solutions crystallized, both vertically (e.g. use of Bash scripts to chain Apache Pig scripts) and horizontally (e.g. shift from Bash usage to Apache Oozie). Taking into consideration all the efforts, we decided to disseminate our experiences in order to recommend the CoAnSys framework, highlight tools available in the Apache Hadoop environment and familiarize readers with the development of research framework.

\section{Tools and Practices}

This section presents some important aspects of a development process and workflow creation tools that are being used in CoAnSys.  

\label{sec:tools_and_practices}
\subsection{Development Aspects}
\label{sec:development_aspects}
Workflows implemented in CoAnSys algorithms \cite{Dendek2013,Matfed:2013} are quite straightforward - no traps, no mazes, very transparent ideas. Crafting them against guidelines mentioned in Section \ref{sec:research_environment} had impact not only on practices used, but also on the choice of tools. CeON team treated with a particular care the following development aspects:
\begin{enumerate}
\item Workflows’ parts have to be loosely coupled.
\item Workflows need to be executed in parallel.
\item Each module should be parametrized.
\item Experiments results have to be maintained by automatic procedure.
\item Experiment options should be separated from execution variables. 
\item Code repetitions should be kept at the lowest possible level.
\item With growth of a workflow, comprehensibility should remain unchanged.
\item Development and production environment options should be separated and easy to maintain.
\item Basic code structures (e.g. loops) have to be provided.
\end{enumerate}
The full application of those guidelines results in a developer experience remote from a na\"{\i}ve implementation.
\subsection{Development Language}
\label{sec:development_languages}
Contrasting the term frequency–inverse document frequency (TF-IDF) calculation code written in Java (418 lines of code) with Apache Pig (28 lines of code) leads to the conclusion that development in Apache Pig may be more productive and transparent. Additionally, incorporating into Pig scripts mechanisms such as the definition of parameters or macros increases their flexibility (point 3). However, due to the lack of loop blocks in Pig, scripts are strongly coupled (points 1 and 9) and weakly reusable. This stresses the need of using gluing scripts. Thus, Bash looms as a natural first aid, especially in a development environment.

Sequences of data transformations described in Pig scripts are broken into the sets of sub-procedures activated by Bash scripts (point 1), which are close to perfect, but a few limitations appear. First of all, some amendments are needed in Bash code to enable fully parallel execution (point 2). Naturally, those mechanisms are not managed by any Hadoop environment load manager.

Furthermore, with each aspect of an experiment, a number of nested loops have been continually growing, leading to a long code that was hard to develop or even comprehend (point 7). Nevertheless, experiment parameters (e.g. the number of neighbour documents) are mixed with auxiliary variables like a fold number (point 5). In addition, managing experiment's results (e.g. choosing localization) have to be done by hand or by additions in bash scripts (point 4).

Development team in CeON was aware that at some point, a general workflow embracing not only document classification model creation, but also all other algorithms has to be prepared and it was quite obvious at that point that Bash is not a convenient solution in this scale.

\subsection{Apache Oozie}
\label{sec:apache_oozie_usage}
While implementing Bash-wise workflow management, we have also started the adaptation of Apache Oozie. That framework outperforms Bash in a few aspects: it is easy to separate parameters from options (point 5) and it allows creation and management of sets or parameters for development or production environment (point 8).

On the other hand, the workflow management system from Apache rises some obstacles: it requires writing long XML files (point 8) (see Lis. \ref{lis:oozie_action_block}), it does not support for loop statements (point 9; simulation via tedious copy-pasting), the output and subsequent executions of one code or its continuously improved versions, have to be manually managed by a developer, and finally, there is no automatic mechanism for creating a link between code versions and the results. 



\lstset{language=XML}
\begin{lstlisting}[caption={The example of an Apache Oozie action block. Each action has (a) its own distinct name, (b) a subsequent nodes for successful (ok) and unsuccessful (error) script finalization, (c) a set of parameters with a path to JAR libraries among others and (d) a script to be executed with.},label={lis:oozie_action_block}]
   <action name='tfidf'>
       <pig>
           <job-tracker>${jobTracker}</job-tracker>
           <name-node>${nameNode}</name-node>
           <prepare>
               <delete path="${ds_similarityOutputPath}"/>
           </prepare>
           <configuration>
               <property>
                   <name>mapred.job.queue.name</name>
                   <value>${queueName}</value>
               </property>
           </configuration>
           <script>${pigScriptsDir}/document-similarity.pig</script>
           <param>inputPath=${ds_bwndataMetadataInputPath}</param>
           <param>outputPath=${ds_similarityOutputPath}</param>
           <param>commonJarPath=${ds_commonJarPath}</param>
           <file>${pigScriptsDir}/macros.pig#macros.pig</file>
       </pig>      
       <ok to='end'/>
       <error to='kill'/>
   </action>  
\end{lstlisting}

\section{Chrum}
\label{sec:chrum}
Alleviation for Apache Oozie drawbacks mentioned in section \ref{sec:apache_oozie_usage} has been introduced in a Python-based tool called Chrum (eng. “oink”) \footnote{\url{http://github.com/CeON/chrum}}. Two main goals of Chrum are as follows:
\begin{enumerate}
	\item Workflow description simplification \begin{itemize}
        \item collapsing repetitive code (like loop statements or configuration information in workflow nodes)
    \end{itemize}
    \item Better experiment management \begin{itemize}
        \item creation of separate spaces for results and code used in a given time
        \item creation of separate spaces for each combination of parameters
		\end{itemize}        
\end{enumerate}

\subsection{Workflow Generation}
\label{sec:chrum_workflow_generation}
Chrum, besides syntax enhancements, introduces additional layer over Apache Oozie configuration and workflow files. In a workflow file, the following new blocks are present:
REPLACE, ACTION and FORK\_MERGE. With these three simple operations, a workflow development process is much more convenient. A scope of a Chrum block is limited by BEG and END keyword, as presented in Lis. \ref{lis:action_block}

\lstset{language=Black}
\begin{lstlisting}[caption={The definition of the ACTION BLOCK docs2neigh\_01. Chrum Manager is going to substitute REPLACE BLOCKS VALUES present inside of the ACTION BLOCK (e.g. WF-1 shown in Lis.\ref{lis:replace_block}) and subsequently expand it to a final form similar to the code presented in Lis. \ref{lis:oozie_action_block}},label={lis:action_block}]
# BEG:ACTION name=docs2neigh_01 ok=createDocClassif_02 error=kill
    @PIG_START@
        @PR-1@
        @CONFIG-1@
        @WF-1@
    @PIG_END@
# END:ACTION
\end{lstlisting}

The ACTION block, presented in Lis. \ref{lis:action_block}, gathers all important Apache Oozie action node information, i.e. a node name and the final node after success or failure. This block may contain other blocks, e.g. replace block variables.

The REPLACE block, shown in Lis. \ref{lis:replace_block}, allows extraction and placing in the same part of a workflow file the essence of node description i.e. parameters, files needed and a script used. In addition, the REPLACE block enables abbreviation of other longer portion of the code by referencing to it with a replace block variable.

\lstset{language=Black}
\begin{lstlisting}[caption={The definition of the REPLACE block WF-1, which contains regular Apache Oozie XML code. A definition of one block may include another Chrum block. The definitions of blocks are not allowed to create a substitution cycle.},label={lis:replace_block}]
# BEG:REPLACE @WF-1@
    <script>${pigScriptsDir}/1_MODEL_CREATE_01_docs2neig.pig</script>
    <param>dc_m_double_sample=${dc_m_double_sample}</param>
	<param>dc_m_hbase_inputDocsData=${dc_m_hbase_inputDocsData}</param>
    <param>dc_m_hdfs_neighs=${dc_m_hdfs_neighs}</param>
    <param>dc_m_int_folds=${dc_m_int_folds}</param>
    @AUXIL@
# END:REPLACE
\end{lstlisting}

Finally, the FORK\_MERGE block, presented in Lis. \ref{lis:fork_merge_block}, enables a user to execute a script in a way similar to a (nested) "for" loop. In the first line of a FORK\_MERGE block one declares the name of a block, accompanied with the names of subsequent nodes in case of success and failure. In the next line(s), parameters with their values separated by spaces are defined, e.g. "@src@ /tmp/1 /tmp/2 /tmp/3". It is possible to dynamically generate the sequence of numeric values using the idiom "seq", e.g. "@fold@ seq(0,\${dc\_m\_int\_folds},1)", where 0 is the initial number of a sequence, \${dc\_m\_int\_folds} is the maximum number of that sequence (defined in the Apache Oozie configuration file) and 1 is a step between consecutive values. Idioms like "seq" can be easily added to Chrum by a developer. 

Eventually, starting from the first line that does not describe parameters, the code for substitution begins. It is duplicated $n$ times, where $n$ is a number of combinations of parameters' values. In each copy of the code, parameters (denoted as @param\_name@) are substituted by corresponding values. Then, each version of the original code is packed into an Apache Oozie node. Calculations in all precedent nodes have to be successfully completed before the execution of a subsequent node begins.

\lstset{language=Black}
\begin{lstlisting}[caption={The definition of the FORK\_MERGE block. FORK\_MERGE block contains the name of a fork node ("split\_03"), the next node after merge ("enrich\_04") and an error node, where control goes in case of any failure.},label={lis:fork_merge_block}]
# BEG:FORK_MERGE name=split_03 node_after_join=enrich_04 error=kill
@src@ ${dc_m_hdfs_neighs} ${dc_m_hdfs_docClassifMapping}
@fold@ seq(0,${dc_m_int_folds},1)
    @PIG_START@
        @PR-3@
        @CONFIG-1@
		<script>${pigScriptsDir}/1_MODEL_CREATE_03_split.pig</script>
		<param>dc_m_hdfs_src=@src@</param>
		<param>dc_m_int_concreteInvestigatedFold=@fold@</param>
		@AUXIL@
    @PIG_END@
# END:FORK_MERGE
\end{lstlisting}

These three simple commands allow to significantly shorten the number of lines of code e.g. for the  document classification model, Chrum shortened an XML workflow file from 1.5 KLOC to 0.1 KLOC. This reduction in size can grow even further when the number of parameters increases. As a result, the clarity of the code is better and a management is easier.

\subsection{Experiments Management}
\label{sec:chrum_experiments_managements}
Frequent inconvenience for researchers and developers is that, even if the code is under version control system, it is hard to follow a bind between the code version and the results. Similarly, when a few combinations of parameters have to be tested against a given code, a control system has to be written or tedious manual actions have to be taken. To alleviate those pains in Hadoop environment, we introduce Chrum Experiment Manager (CEM). CEM, as an input, takes Apache Oozie workflow and properties files, accompanied with a Chrum configuration file. The following information is specified in the Chrum configuration file:
\begin{enumerate}
\item Key information \begin{enumerate}
	\item Project name
	\item HDFS localization of data specified
	\item Local path to Chrum trigger scripts 
	\item Apache Oozie server address and port.
	\end{enumerate}
\item Folders to be created in the \$PROJECT directory with a content indicated (e.g. "lib $\leftarrow$ /usr/lib/pig/pig-0.9.2-cdh4.0.1.jar")
\end{enumerate}
CEM execution triggers key information extraction from the Chrum configuration.
For each combination of multivalued properties in Apache Oozie properties file, a modified copy of Apache Oozie properties file is created on a local file system, each of which carries CEM execution time (in the property "COMPILATION\_TIME") and properties combination ("PARAMETER\_COMBINATION"). Multivalued properties (denoted as "@var@ val1 val2 val3") are flattened to one variable (e.g. "var=val2"), where each combination appears once. 
Properties files composed in such a manner accompanied with generated workflow files and workflow submission script (execute-in-oozie.py) are stored in the local file system (\$LOCAL/\$PROJECT/\$COMPILATION\_TIME/\\\$PARAMETER\_COMBINATION), while libraries, scripts, workflow files 
and other data mentioned in the Chrum configuration file are sent to the HDFS folders \\ \$HDFS/\$PROJECT/\$COMPILATION\_TIME/\$PARAMETER\_COMBINATION. Finally, when execute-in-oozie.py script is started the current time is stored in a propertie file ("EXECUTION\_TIME"), followed by a workflow submission to Apache Oozie Server.

Thanks to this procedure, the following goals can be achieved:
\begin{enumerate}
\item A user, in an easy way, can send to HDFS all data (e.g. source code, personal notes) needed for further consideration. 
\item Each call to CEM and each combination of parameters are stored in a different localization (\$HDFS/\$PROJECT/\$COMPILATION\_TIME/\\\$PARAMETER\_COMBINATION).
\item Each workflow execution output may be stored in a separate localization for subsequent investigation \\ (\$HDFS/\$PROJECT/\$COMPILATION\_TIME/\$PARAMETER\_COMBINATION/\\results/\$EXECUTION\_TIME).
\end{enumerate}

\subsection{Comparison of Workflow Creation and Management Approaches}
\label{sec:comparison_workflow_creation}

Contributing to CoAnSys as well as seeking for the most clear and coherent code has been a path from using only Apache Pig scripts, mixture of Apache Pig and Bash, Apache Oozie and finally Chrum. Workflows described with those approaches may be analyzed in terms of the following aspects:
\begin{enumerate}
\item Strong code coupling.
\item Loop statement support.
\item Separation of options and parameters.
\item Cost of code maintenance - both in micro and macro scale.
\item Extensibility.
\item Solution maturity.
\end{enumerate}
Ours subjective opinion, reflected in Tab. \ref{tab:workflows}, is that mixture of Apache Pig and Chrum gives the best circumstances for developing sophisticated algorithms, however it has to be stressed that this solution is both immature and implicitly dependent on other tools, like Apache Oozie. Chrum gives also a way to easily extend its capabilities by introducing new functions, similar to \textit{seq} function mentioned in Sec. \ref{sec:chrum_workflow_generation}.

\begin{table}
\caption{Subjective assessment of different approaches to a workflow creation.}
\begin{tabular}{|>{\centering}m{1.3cm}|>{\centering}m{1.2cm}|>{\centering}m{1.0cm}|>{\centering}m{1.8cm}|>{\centering}m{1.8cm}|>{\centering}m{1.2cm}|>{\centering}m{1.3cm}|>{\centering}m{1.5cm}|}
\hline 
Approach & Strong code coupling & Loops support & Cost of code maintenance in small scale & Cost of code maintenance in large scale & Extensi-bility & Maturity & Option / parameters separation\tabularnewline
\hline 
\hline 
Pig & high & none & high & very high & low & high & low\tabularnewline
\hline 
Bash & low & present & moderate & high & high & high & moderate\tabularnewline
\hline 
Oozie & low & none & low & high & low & moderate & moderate\tabularnewline
\hline 
Chrum & low & present & low & moderate & high & low & high\tabularnewline
\hline 
\end{tabular}
\label{tab:workflows}
\end{table}

\section{Summary and Future Work}
\label{sec:future_works}
Chrum is still far from its maturity and as so, there is plenty of room for improvements. Proposed transformations are the most frequent, but there are still a few Apache Oozie operations which could be transformed into standard procedural language statements like while, if, etc. As Chrum enables an easy way to store subsequent versions of the code and its output, it would be beneficial to develop a convenient way of comparing results in terms of a given metric.

In this article we have described an effective approach to conducting research in the Apache Hadoop environment, based on presented Content Analysis Framework and the analytic stack used in CeON. We have shown obstacles emerging in a development and a research process and provided suitable solutions which may have positive effect on both of them. 

We hope that the provided description of our experience gained from the usage of Apache Hadoop will lower the entry barrier for users and researchers who want to benefit from MapReduce paradigm and it will reduce inconveniences arising in the process of learning a new technology.

\bibliographystyle{splncs}
\bibliography{bibliography}
 
\end{document}